\RequirePackage{fix-cm}
\documentclass[twocolumn]{svjour3}          
\smartqed  
\usepackage{graphicx}
\usepackage{subfigure}
\usepackage[colorlinks]{hyperref}
\usepackage{ctable} 
\usepackage{adjustbox}

\begin{document}

\providecommand{\pasa}{Publ.\ Astron.\ Soc.\ Austr.}
\providecommand{\aap}{Astron.\ Astrophys.}
\providecommand{\apjs}{Astrophys.\ J.\ Suppl.}
\providecommand{\apjl}{Astrophys.\ J.}
\providecommand{\apj}{Astrophys.\ J.}
\providecommand{\physrep}{Phys.\ Rep.}
\providecommand{\mnras}{Mon.\ Not.\ Roy.\ Astron.\ Soc.}
\providecommand{\npa}{Nucl.\ Phys.\ A}
\providecommand{\plb}{Phys.\ Lett. \ B}
\providecommand{\epj}{Eur.\ Phys.\ J.}
\providecommand{\epja}{Eur.\ Phys.\ J.\ A}
\providecommand{\rmp}{Rev/\ Mod.\ Phys.}
\providecommand{\prc}{Phys.\ Rev.\ C}
\providecommand{\prd}{Phys.\ Rev.\ D}
\providecommand{\pr}{Phys.\ Rev.}
\providecommand{\prl}{Phys.\ Rev.\ Lett.}
\providecommand{\npa}{Nucl.\ Rev.\ A}
\providecommand{\nat}{Nature}

\title{QCD phase transition drives supernova explosion of a very massive star\\\vspace{-5mm}}
\subtitle{Dependence on metallicity of the progenitor star}

\titlerunning{Eur. Phys. J. A}        

\author{Tobias Fischer}

\authorrunning{Eur. Phys. J. A} 
\institute{
University of Wroclaw, Plac M. Borna 9, 50-204 Wroclaw, Poland \\
\email{tobias.fischer@uwr.edu.pl} 
}

\date{Received: 28~May~2021 / Accepted: 17~August~2021\\
\copyright The Author(s)~2021\\
Communicated by Laura Tolos
}

\maketitle

\begin{abstract}
The nature of core-collapse supernova (SN) explosions is yet incompletely understood. The present article revisits the scenario in which the release of latent heat due to a first-order phase transition, from normal nuclear matter to the quark--gluon plasma, liberates the necessary energy to explain observed SN explosions. Here, the role of the metallicity of the stellar progenitor is investigated, comparing a solar metallicity and a low-metallicity case, both having a zero-age main sequence (ZAMS) mass of 75~M$_\odot$. It is found that low-metallicity models belong exclusively to the failed SN branch, featuring the formation of black holes without explosions. It excludes this class of massive star explosions as possible site for the nucleosynthesis of heavy elements at extremely low metallicity, usually associated with the early universe.
\end{abstract}

\section{Introduction}
\label{sec:intro}
All stars more massive than about 8 times the mass of the sun (M$_\odot$) end their life as a core-collapse SN. These events belong to the most powerful outbursts in the modern universe, where most of the energy is carried in form of neutrinos. SN explosions are associated with the revival of the stalled bounce shock. The latter forms when the collapsing stellar of a massive star reaches central densities slightly in excess of normal nuclear matter density ($\rho_{\rm sat}\simeq 2.5\times 10^{14}$~g~cm$^{-3}$, equivalent $0.15$~fm$^{-3}$ in nuclear units), when the short-range repulsive nuclear force balances gravity such that the collapsing core bounces back. The extreme conditions reached at the SN interior, supersaturation densities and temperatures on the order of 10--50~MeV as well as large neutron excess given by the electron abundance of $Y_e=0.01-0.3$, make the core-collapse SN ideal high-energy nuclear and particle physics laboratories~\cite{Fischer:2017}.

Several SN explosions scenarios, i.e. the liberation of energy from the nascent compact central hot and dense object---the proto-neutron star (PNS)---to the bounce shock have been proposed; the magneto-rotational mechanism~\cite{LeBlanc:1970kg} and the presently considered {\em standard} neutrino-heating mechanism~\cite{Bethe:1985ux}. The latter, however, requires multi-dimensional simulations where the neutrino heating efficiency increases in the presence of convection and hydrodynamics instabilities, to obtain massive star explosions for iron-core progenitors more massive than about 9~M$_\odot$~\cite{Janka:2007hi,Mirizzi:2016}. The exceptions are the slightly less massive progenitors in the zero-age main sequence mass range of 8--9~M$_\odot$~\cite{Nomoto:1987,Kitaura:2006,Melson:2015}, for which neutrino-driven SN explosions can be obtained in spherical symmetric simulations. 

The present article focuses on an additional SN explosion mechanism. It is related to the energy, more precisely latent heat release, due to a first-order phase transition from normal nuclear matter to the quark--gluon plasma at high density~\cite{Takahara:1988yd,Gentile:1993ma,Sagert:2008ka}. First principle calculations of Quantum Chromodynamics (QCD), the theory of strong interactions with quarks and gluons as the degrees of freedom, are only possible at vanishing baryon densities, where the QCD equations are solved numerically being mapped onto a space-time lattice. These computations predict a smooth cross-over transition at a pseudo-critical temperature in the range of 150--160~MeV~\cite{Bazavov:2014,Borsanyi:2014,Bazavov:2019}. At finite and high baryon density, phenomenological quark matter models have commonly been employed in astrophysical studies of compact stellar objects. 

The classical but simplistic quark-bag model EOS~\cite{Farhi:1984qu} has been the foundation in the core-collapse SN study of Ref.~\cite{Sagert:2008ka}. It is based on the equation of state (EOS) of non-interacting quarks, shifted by a constant bag pressure. One of the most striking problems for this class of quark-matter hybrid EOS in applications to astrophysical studies, featuring hot and dense matter, is the difficulty to obtain maximum neutron star masses compatible with the current observational constraint of about 2~M$_\odot$~\cite{Antoniadis:2013,Fonseca:2016,Cromartie:2020NatAs,Fonseca:2021}. This caveat has been overcome recently, with the introduction of repulsive quark matter interactions~\cite{Benic:2015,Klaehn:2015,Kaltenborn:2017}, providing sufficient stiffness with increasing density. The extension to finite temperatures and arbitrary isospin asymmetry is seemingly trivial for the class of quark-bag model EOS~\cite{Klaehn:2017}. It represents a challenge for microscopic quark-matter EOS, such as the Nambu--Jona-Lasinio models~\cite{NJL:1961,Buballa:2003qv,Klaehn:2013} as well as Schwinger--Dyson models~\cite{Roberts:2000aa,Chen:2008zr,Chen:2011my,Bashir:2012fs,Cloet:2013jya,Chen:2015mda}. These phenomenological approaches are applicable to the non-perturbative regime of QCD, however, they are built on a certain truncation of QCD. Most importantly, they are lacking confinement. On the other hand, perturbative QCD is valid only near the limit of asymptotic freedom, where quarks are no more strongly coupled~\cite{Kurkela:2009gj,Kurkela:2014vha,Kurkela:2020NatPh}. 

The recently developed phenomenological {\em string-flip} quark-matter EOS is the foundation of the present work. It features not only a first-order phase transition from normal nuclear matter, it is also built on a mechanism for (de)confinement~\cite{Kaltenborn:2017}. It has been extended to arbitrary isospin asymmetry and finite temperatures~\cite{Fischer:2018,Bastian:2021}. With model parameters selected such that the conditions for the hadron--quark phase transition to occur at the interior of massive compact stars, i.e. densities for the onset of quark-matter on the order of 2--5~$\times\rho_{\rm sat}$, one is led to the conclusions that the associated SN explosion mechanism is likely to operate for the high-mass end of core-collapse SN progenitors, associated with the ZAMS mass of $>30$~M$_\odot$~\cite{Fischer:2018}, whereas less massive progenitor stars would be subject to other SN explosion scenarios. Note that also the failed SN branch, in which a black hole forms instead of an explosion, is considered within the class of hadron--quark phase transition SN. It depends on the compactness, more precisely, on the mass enclosed inside the PNS at the onset of the hadron--quark phase transition. If the latter exceeds the maximum mass of the hadron--quark hybrid EOS, then an explosion is likely to fail. 

The string-flip class of hadron--quark hybrid EOS has also been the foundation for studies of binary neutron star mergers~\cite{Bauswein:2019PhRvL,Bauswein:2020PhRvL,Blacker:2020}. One of the key goals has been the identification of a first-order phase transition through possibly observable signatures, in the neutrino signal from SN and gravitational waves~\cite{OConnor:2020}, the latter also in the context of binary neutron star mergers~\cite{Most:2019PhRvL}.

One important aspect of the SN explosion scenario driven by the QCE phase transition is the role of metallicity\footnote{Astronomers define as metals all elements heavier than helium.}, which has not yet been addressed. It strongly affects the evolution of massive stars through mass loss at the star's surface occurring during the different nuclear burning stages, and hence the core structure at the onset of core collapse. This article reports the SN explosion of a very massive ZAMS progenitor star of 75~M$_\odot$ of solar metallicity, driven by the hadron--quark phase transition, in comparison with those launched from a 75~M$_\odot$ of extremely low metallicity. 

The manuscript is organised as follows: In Sect.~\ref{sec:eos} the hadron--quark matter hybrid EOS is revisited. Section~\ref{sec:SN} reviews briefly the SN model employed in this study together with a discussion of the stellar progenitors. The SN simulations are discussed in Sect.~\ref{sec:results}, and the manuscript closes with the summary and conclusions in Sect.~\ref{sec:sum}.

\section{Dense matter equation of state with phase transition}
\label{sec:eos}
For the present study the reference hadronic DD2F EOS is implemented. It is a nuclear relativistic mean-field (RMF) model with density-dependent meson--nucleon couplings~\cite{Typel:2005,Typel:2009sy}, which has been modified from the original DD2 RMF parametrization in order to be consistent with the flow constraint~\cite{Danielewicz:2002Sci}. As a consequence, DD2F is slightly softer than DD2 at densities in excess of saturation density. At low densities, below the saturation density, and at temperatures below about 15--20~MeV, nuclear clusters are taken into account based on the nuclear statistical equilibrium model of Ref.~\cite{Hempel:2009mc}, with several 1000 nuclear species with tabulated and partly calculated nuclear masses. It implements the dissolving of all clusters through a geometric excluded volume approach, which can be adjusted to calculations of in-medium nuclear properties in order to overcome the limitations of the statistical model~\cite{Fischer:2020c}. 

\begin{table*}
\centering
\caption{DD2F-RDF~1.2 hybrid EOS properties, onset density for the phase transition ($\rho_{\rm onset}$), the density for reaching the pure quark-matter phase ($\rho_{\rm final}$), as well as inset neutron star mass ($M_{\rm onset}$) and maximum mass ($M_{\rm max}$) together with the central density of the maximum mass configuration ($\rho_{\rm central}\vert_{M_{\rm max}}$), for $T=0$ and a constant entropy per particle of $s=3~k_{\rm B}$.}
\begin{tabular}{l cc c cc c cc c cc c cc c cc c cc c cc c}
\hline
DD2F-RDF~1.2 & && $\rho_{\rm onset}$ & && $\rho_{\rm final}$ & && $M_{\rm onset}$ & && $M_{\rm max}^{\rm RDF~1.2}$ & && $\left.\rho_{\rm central}\right\vert_{M_{\rm max}}$ & && $M_{\rm max}^{\rm DD2F}$ \\
 & && $[\rho_{\rm sat}]$ & && $[\rho_{\rm sat}]$ & && $[{\rm M}_\odot]$ & && $[{\rm M}_\odot]$  & && $[\rho_{\rm sat}]$ & && $[\rho_{\rm sat}]$ \\
\specialrule{.1em}{.05em}{.05em} \\
\vspace{1mm}
$T = 0$ & && 3.05 &  && 3.60 & && 1.37 & && 2.16 & && 7.14 & && 2.09 \\
\vspace{1mm}
$s = 3~k_{\rm B}$ & && 1.15 & && 3.25 & && 1.44 & && 2.07 & && 7.22 & && 2.17 \\
\hline
\end{tabular}
\label{tab:eos}
\end{table*}

The DD2F hadronic EOS has been extended to take into account a phase transition to the quark--gluon plasma. Therefore, the string-flip EOS for quark matter is employed~\cite{Kaltenborn:2017}. It is based on the relativistic density functional (RDF) approach to quark matter. The present work uses the parametrization denoted as RDF~1.2 from the comprehensive catalog of Ref.~\cite{Bastian:2021}\footnote{All EOS of the present study are available at the CompOSE data base under https://compose.obspm.fr~\cite{Typel:2014}.}.  For the phase-transition construction between the DD2F hadronic EOS and the string-flip EOS, Maxwell's condition of phase equilibrium has been applied, i.e kinetic equilibrium for the pressures with respect to the baryon chemical potential, for fixed temperature, $T$, and charge chemical potential, $\mu_Q$. The latter is defined as the difference between proton and neutron chemical potentials, $\mu_p$ and $\mu_n$, $\mu_Q=\mu_p-\mu_n$ in the hadronic phase, and as the difference between up and down chemical potentials, $\mu_u$ and $\mu_d$, $\mu_Q=\mu_u-\mu_d$ in the quark matter phase. Maxwell's construction results in a jump condition for all thermodynamic quantities, from the pure hadronic phase to the pure quark matter phase. However, for practical purposes in applications of astrophysical simulations, data are needed in the coexistence region in between. Therefore, a quark-volume fraction has been employed, with a linear dependence on the baryon density. Further details can be found in Ref.~\cite{Bastian:2021}.

\begin{figure*}
\centering
\vspace{2mm}
\subfigure[~$T=0$]{
\includegraphics[width=0.875\textwidth]{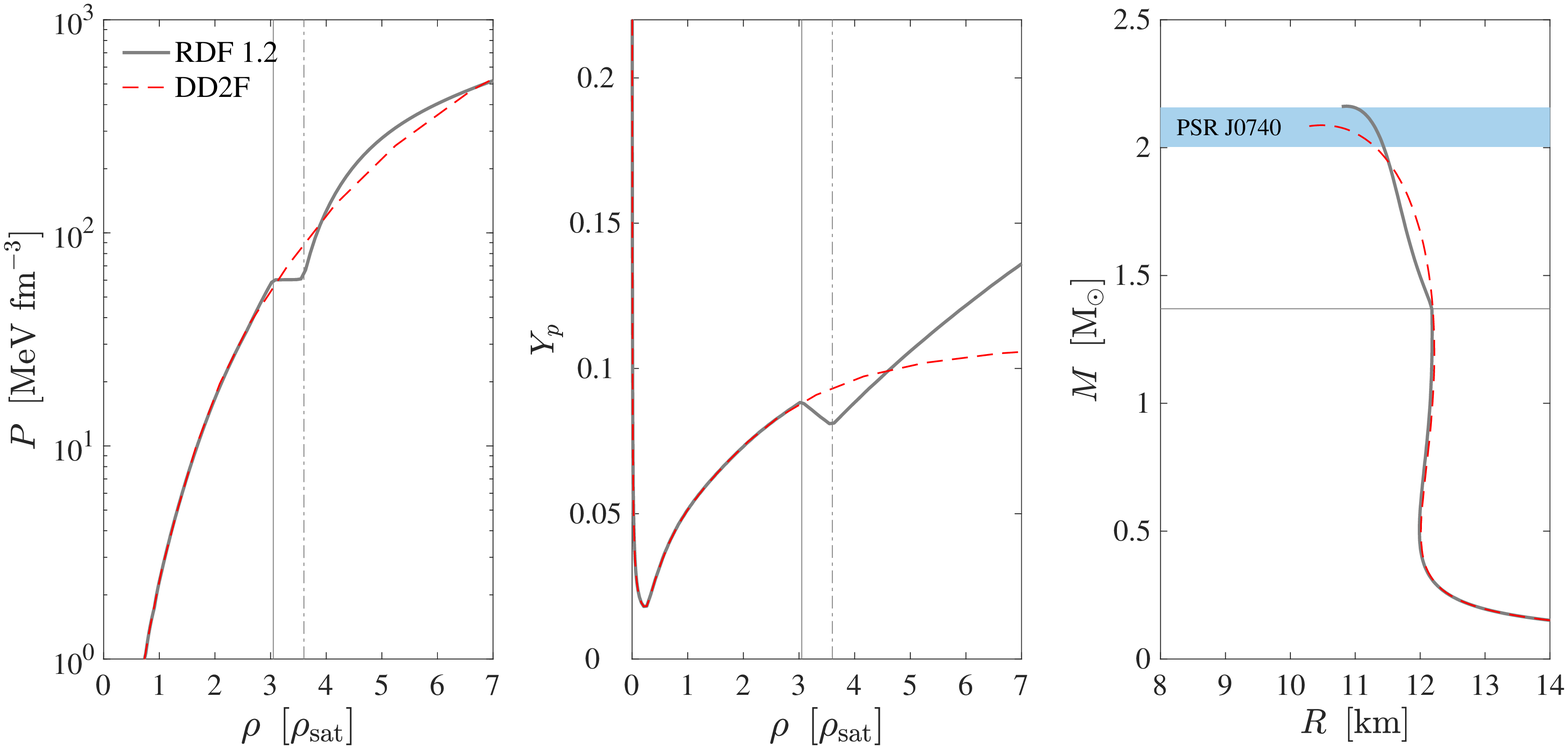}
\label{fig:eos_a}}
\\
\subfigure[~$s=3~k_{\rm B}$]{
\includegraphics[width=0.875\textwidth]{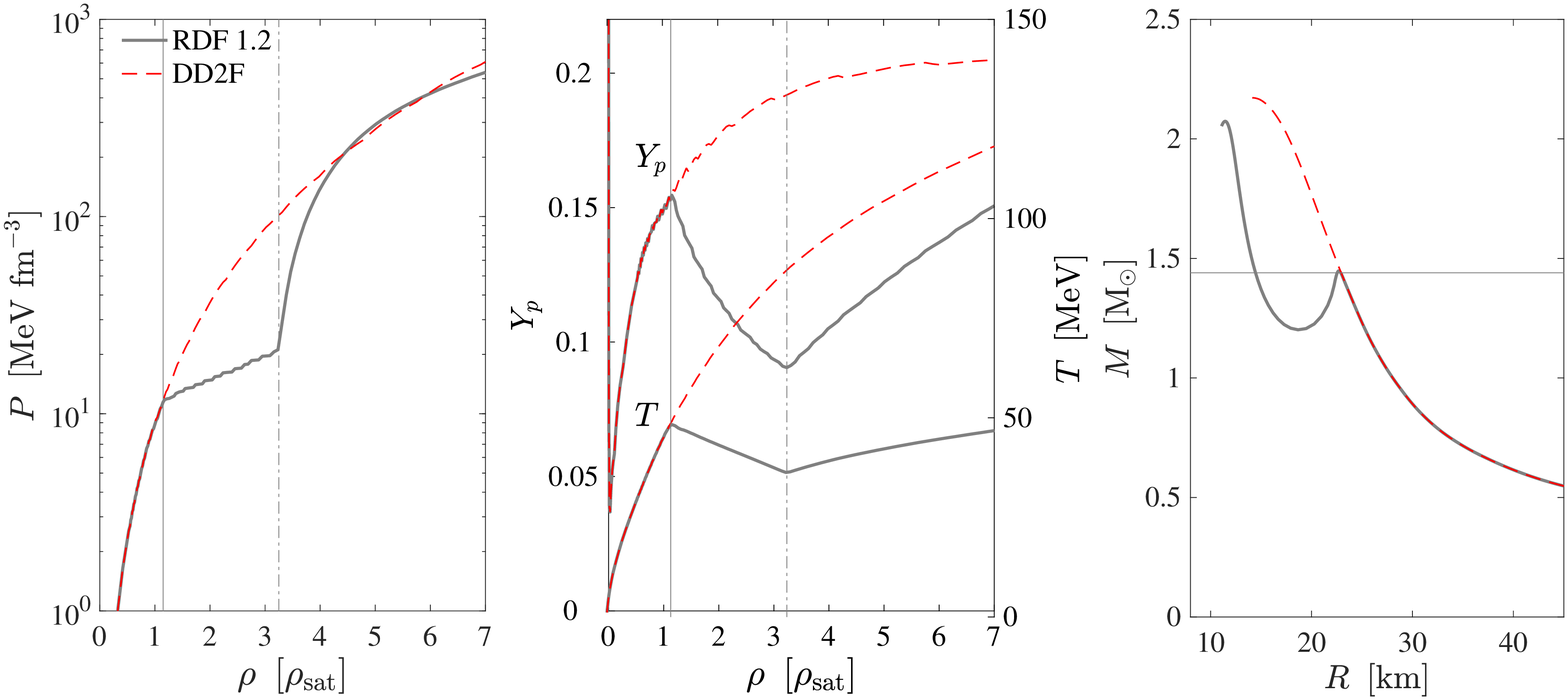}
\label{fig:eos_b}}
\caption{EOS under the condition of $\beta$-equilibrium, showing the pressure $P$, the proton abundance $Y_p$ and the temperature $T$ (only for finite entropy), with respect to the restmass density $\rho$, for $T=0$ in graph~(a) and constant entropy per particle of $s=3~k_{\rm B}$ in graph~(b), as well as the corresponding mass--radius relations, comparing the hadronic DD2F reference model (red dached lines) and SF2 (solid gray lines). The thin vertical solid and dash-dotted lines mark the location for the onset of the phase transition and when reaching the pure quark-matter phase, respectively, whereas the horizontal line in the mass--radius relation graphs marks the corresponding onset mass, together with the present maximum neutron star constraint of $2.08\pm 0.07$~M$_\odot$ from the measurement of the pulsar PSR~J0740+6620~\cite{Antoniadis:2013,Fonseca:2021}.}
\label{fig:eos}
\end{figure*}

The class of string-flip EOS is build on a confining mechanism through divergent particle masses towards vanishing density. It still remains to be shown in how far it matches the (dynamical) restoration of chiral symmetry. In addition, in order to obtain sufficient stiffness at high densities, repulsive interactions are taken into account. The original string-flip EOS includes besides linear terms  higher-order repulsive interactions, which are omitted here in the RDF~1.2 version, such that the problem of causality---when the speed of sound approaches the speed of light---does not occur in the density range encountered at the interior of compact stellar objects. The string-flip model has been extended to include an isospin mean field, i.e. the vector–isovector interaction channel, which are comparable to the $\rho$-meson interactions in nuclear RMF-type models. The coupling can be adjusted to a smooth behavior of the symmetry energy across the phase transition boundary (a detailed discussion of this aspect is given in Sect.~II~C3. of Ref.~\cite{Bastian:2021}).

Selected properties of the resulting DD2F-RMF~1.2 hybrid EOS are listed in Table~\ref{tab:eos}, for zero temperature and finite entropy per particle of $s=3~k_{\rm B}$. Particularly interesting is the reduction of the onset density for the hadron--quark phase transition, $\rho_{\rm onset}$, with increasing entropy, viz. temperature, while the final density for reaching the pure quark matter phase, $\rho_{\rm final}$, remains nearly constant for the temperatures of relevance here on the order of $T=0-50$~MeV. These aspects will become relevant in the SN simulations which will be discussed below.

The DD2F-RDF~1.2 hadron--quark hybrid EOS in $\beta$-equilibrium is illustrated in Fig.~\ref{fig:eos}a for $T=0$ and in Fig.~\ref{fig:eos}b for finite entropy per particle of $s=3\, k_{\rm B}$, in comparison to the hadronic DD2F reference EOS. Note that the pressure, $P$, contains all contributions, hadrons and/or quarks as well as electrons, positrons and photons. The last two are added following Ref.~\cite{Timmes:1999}. The vertical lines mark $\rho_{\rm onset}$ (solid lines) and $\rho_{\rm final}$ (dash-dotted lines), where it is interesting to note that while for $T=0$ the pressure is nearly constant in the hadron--quark mixed region, for finite entropy case there is a finite slope. The latter aspect is attributed to the fact that there are different temperatures for the phase transition, $T(\rho_{\rm onset})$, and for reaching the pure quark-matter phase, $T(\rho_{\rm final})$. In order to preserve the entropy per particle, the temperature decreases throughout the mixed phase, for entropies less than about $s\simeq 6-7~k_{\rm B}$, while the temperature increases for higher entropies. This feature is already illustrated in Fig.~1 in Ref.~\cite{Fischer:2018} (see also the solid green lines in Fig.~\ref{fig:phasediagram} which will be discussed below).

\begin{figure*}
\adjustbox{valign=t}{%
\begin{minipage}[t]{.25\linewidth}
\caption{Comparison of the progenitors with a ZAMS masses of 75~M$_\odot$ but different metallicities, from Ref.~\cite{Woosley:2002zz}, showing the density $\rho$ and entropy per particle $s$ (left panels) and the abundance of helium denoted as He (dotted lines), carbon and oxygen denoted as C+O (thin dash-dotted lines) as well as silicon and sulphur denoted as Si+S (dashed lines) and iron-group nuclei collectively denoted as 'Fe' (thick solid lines), with respect to the mass coordinate. In addition, for the low-metallicity progenitor, $^{56}$Ni is shown as thin solid line.\label{fig:progenitor}}
\end{minipage}}%
\hspace{5mm}
\adjustbox{valign=t}{%
\begin{minipage}[t]{0.75\linewidth}
  \begin{subfigure} 
  \centering
  \includegraphics[width=0.975\textwidth]{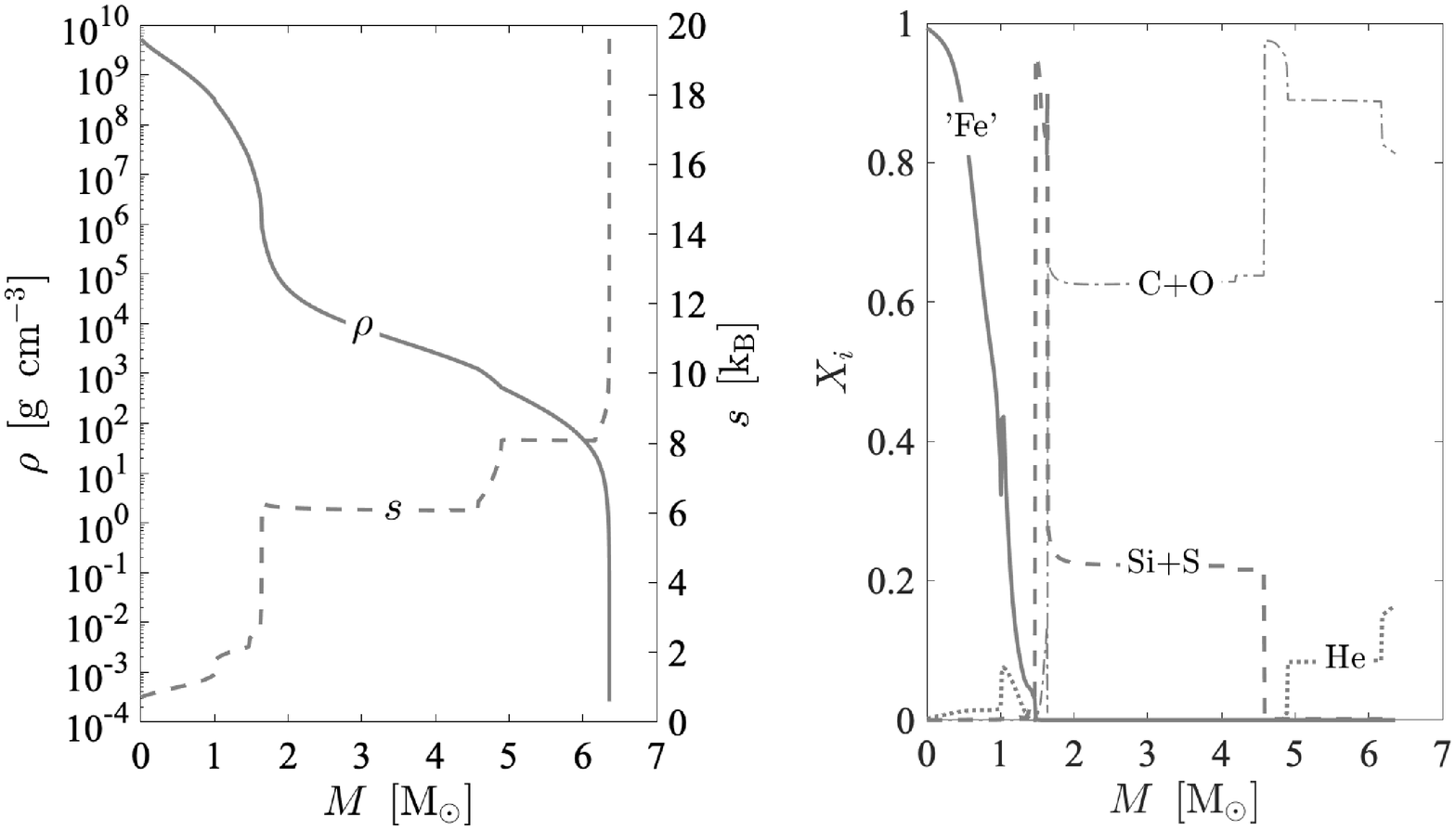} \\
  \centerline{{\bf (a)}~Solar metallicity} \\
  \end{subfigure}\par\bigskip
  \begin{subfigure} 
  \centering
  \includegraphics[width=0.975\textwidth]{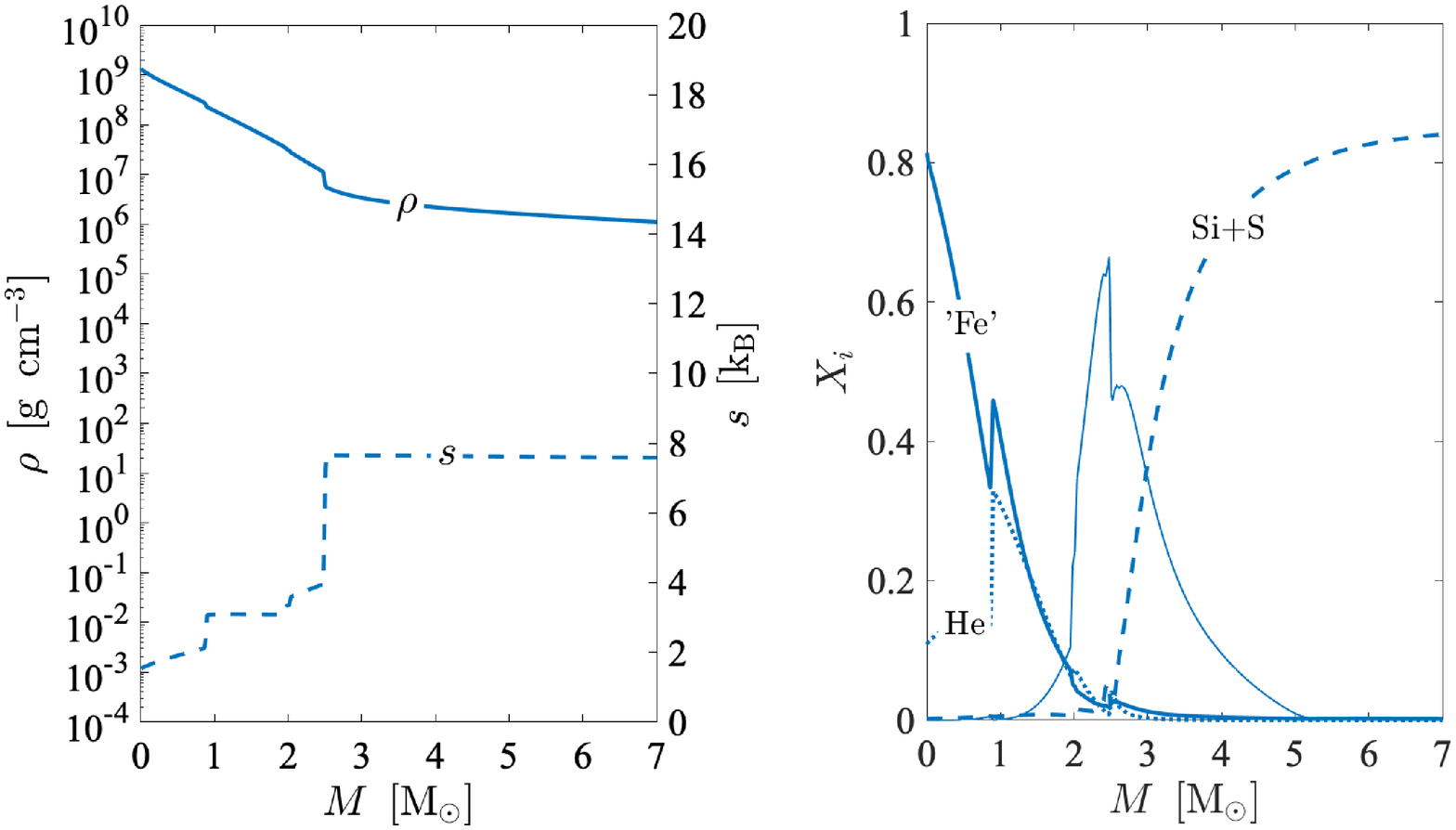} \\
  \centerline{{\bf (b)}~$Z=10^{-4}\times Z_\odot$}
  \end{subfigure}\par\bigskip
\end{minipage}
}
\end{figure*}

It is also interesting to note that the aforementioned Maxwell construction for the phase transition results in gradients for various quantities, e.g., the proton fraction, in general the charge fraction across the coexistence region, in order to conserve the charge chemical potential (see the middle panels in Fig.~\ref{fig:eos}). This behavior is imposed by the linear dependence on density in the coexistence region. For a {\em real} Maxwell construction, both pressure and charge fraction would jump from $\rho_{\rm onset}$ to $\rho_{\rm final}$. Note that in quark matter we construct proton and neutron abundances, $Y_p$ and $Y_n$, from the abundances of up- and down-quark, $Y_u$ and $Y_d$, as follows: $Y_p=2Y_u+Y_d$ and $Y_n=Y_u+2Y_d$.

The corresponding mass--radius relations are shown the right panels of Fig.~\ref{fig:eos}. For the DD2F-RDF~1.2 parameters, at $T=0$, the maximum mass of the hybrid star exceeds the maximum mass of the DD2F reference hadronic EOS; the exact values are given in Table~\ref{tab:eos}. The onset mass for the hybrid branch, i.e. the third family of compact stellar objects, is at $M_{\rm onset}=1.37$~M$_\odot$. These hydrostatic properties change when considering hot configurations, in particular at entropy of $s=3~k_{\rm B}$, which represents a canonical value at the PNS interior. Note that it has been found that the universal relations~\cite{YagiYunes:2013Sci,YagiYunes:2013PhRvD,Maselli:2013} hold only for constant but low values of the entropy~\cite{Raduta:2020,Sedrakian:2021}, since the entropy is a conserved quantity. Since the latter is also the case for SN, it is suitable to consider the situation of constant entropy, rather than constant temperature, in order to understand what happens in simulations of core-collapse SN regarding thermal effects. The first observation is that the coexistence region extends due to the different behaviors of hadronic and quark matter EOS with increasing temperature (see Fig.~\ref{fig:eos}b), besides the significantly larger gradients between $\rho_{\rm onset}$ (vertical solid lines) and $\rho_{\rm final}$ (vertical dash-dotted lines). Secondly, the radii of the hydrostatic configurations become significantly larger, on the order of 20--40~km. Secondly, while the maximum mass of the DD2F reference hadronic EOS increases, the maximum mass of the DD2F-RDF~1.2 hybrid configuration reduces. Moreover, an extended region of gravitational instability develops after the onset of quark matter at the transition from the stable hadronic to the stable quark matter branch~\cite{Hempel:2016}. This {\em twin} phenomenon has long been studied in the context of cold neutron stars, i.e. the existence of two compact stars with the same gravitational mass but different radii and hence belonging to different families of compact stars (cf. Ref.~\cite{Benic:2015} and references therein). It is interesting to note that the twin phenomenon is only present in case of finite, and in particular high temperatures, illustrated by the temperature jump around $T\simeq 50$~MeV at the phase transition (middle panel in Fig.~\ref{fig:eos}b). 

\begin{figure}
\centering
\includegraphics[width=0.5\textwidth]{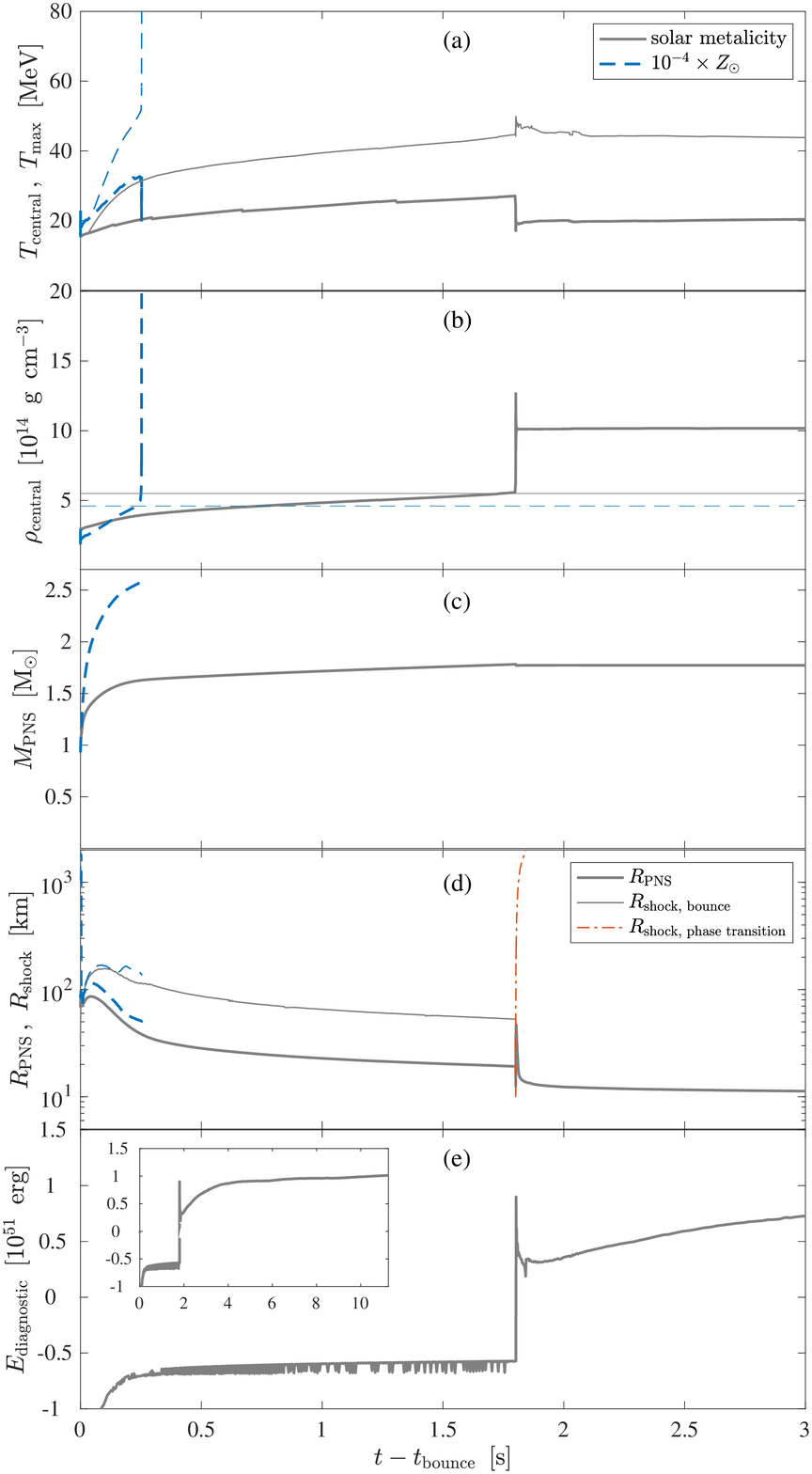}
\caption{Post-bounce evolution of the central and maximum temperatures in graph~(a), $T_{\rm central}$ (thick lines) and $T_{\rm max}$ (thin lines), the central density $\rho_{\rm central}$ in graph~(b) and the PNS mass $M_{\rm PNS}$ in graph~(c) as well as in graph~(d) the PNS radius $R_{\rm PNS}$ (thick lines) and the SN shock radius $R_{\rm shock}$ (thin lines), including the second shock wave due to the hadron--quark phase transition (red dash-dotted lines), as well as in graph~(e) the diagnostic energy $E_{\rm diagnostic}$, comparing the SN simulations of the 75~M$_\odot$ progenitors of solar metallicity (solid gray lines) and $10^{-4}$ of solar metallicity (blue dashed lines). The thin horizontal lines in graph~(b) mark the onset densities for the hadron--quark phase transition for the corresponding temperatures and a represented proton fraction of $Y_p=0.3$.}
\label{fig:evolution}
\end{figure}

\section{Supernova model and stellar progenitors}
\label{sec:SN}
The present numerical study of core-collapse SN is based on {\tt AGILE}{\tt-BOLTZTRAN} (see Ref.~\cite{Liebendoerfer:2004} and references therein), which is a general relativistic neutrino-radiation hydrodynamics model in spherical symmetry featuring accurate 6-species Boltzmann neutrino transport~\cite{Fischer:2020d}. The set of non-muonic weak interactions considered here can be found in Table~1 of Ref.~\cite{Fischer:2020a}, with the references given accordingly, and the muonic weak processes are introduced and discussed in Refs.~\cite{Guo:2020,Fischer:2020d}. {\tt AGILE-BOLTZTRAN} has an adaptive baryon mass mesh refinement~\cite{Liebendoerfer:2002,Fischer:2009af}. In the present simulations 207 radial grid points are used, and the neutrino phase space is discretized following Refs.~\cite{Bruenn:1985en,Mezzacappa:1993gm}, with 6 neutrino lateral angle bins, $\cos\theta\in\{-1,+1\}$, and 36 neutrino energy bins, from 0.5--300~MeV.

For the present investigation of the phase transition in core-collapse SN as explosion mechanism, one of the yet-incompletely addressed questions is the dependence of the stellar progenitor model. In particular, what is the maximum progenitor mass that the hadron--quark phase transition is able to drive the explosion, and how does it depend on metallicity, denoted as $Z$? Therefore, the most massive stellar progenitor is selected from the comprehensive stellar evolution series of Ref.~\cite{Woosley:2002zz}, with a ZAMS mass of 75~M$_\odot$, while a stellar model will ZAMS mass of 50~M$_\odot$ has been explored previously~\cite{Umeda:2007wk,Fischer:2018}.

Radial profiles, as a function of the enclosed mass, for selected quantities are shown in Fig.~\ref{fig:progenitor}a for the solar metallicity model. The current value of the solar surface metallicity is $Z_\odot=0.0187\pm0.0009$, \cite{Asplund:2009,Asplund:2021}. This features a rather compact stellar core at the onset of stellar collapse, with an iron-core mass of about 1.5~M$_\odot$ and with central densities in excess of $5\times 10^9$~g~cm$^{-3}$, as well as a narrow silicon--sulfur shell above the iron core (dashed line in the right panel of Fig.~\ref{fig:progenitor}a. Typical for solar-metallicity stars is a large mass-loss rate, which results in a sharp transition from the stellar core, i.e. carbon-oxygen shell (dash-dotted line in the right panel of Fig.~\ref{fig:progenitor}a) and the transition to the helium-rich envelope, with a sharp density drop over many orders of magnitude. The resulting total enclosed mass is about 6.36~M$_\odot$ at the onset of stellar core collapse~\cite{Woosley:2002zz}, i.e. about 68.64~M$_\odot$ has been lost during the previous nuclear burning stages, mostly during hydrogen and helium burning~\cite{Woosley:2002zz}. The later collision of the SN ejecta with the material ejected pre-collapse may power the brightest SN explosions known as super-luminous SN~\cite{Fischer:2018}.

The situation is very different for stars with lower metallicity. Figure~\ref{fig:progenitor}b shows the same radial profiles for the same 75~M$_\odot$ ZAMS mass but at a metallicity of $10^{-4}$ of the solar value, from the same stellar evolution series~\cite{Woosley:2002zz}. This stellar model has experienced nearly zero mass loss. It it related with the photon opacity, which scales with the number of protons and is hence substantially higher for ions. The lack of efficient mass loss leaves a nearly as massive progenitor star, with a total mass at the onset of stellar collapse of 75~M$_\odot$~\cite{Woosley:2002zz}, featuring a central density of about $10^9$~g~cm$^{-3}$ and a higher central entropy per baryon of about $s\simeq 2~k_{\rm B}$, in comparison to the solar metalicty progenitor with $s\simeq 1.5~k_{\rm B}$ (see Fig.~\ref{fig:progenitor}). Note also that the nascent iron core contains nearly 3~M$_\odot$ (the abundance of $^{56}$Ni is shown as thin solid line in the right panel of Fig.~\ref{fig:progenitor}b). 

\section{Supernova simulations}
\label{sec:results}
In this section, results will be discussed from simulations of core-collapse SN, launched from the stellar progenitors introduced in Sect.~\ref{sec:SN}, with a ZAMS mass of 75~M$_\odot$ of solar metallicity and $10^{-4}$ of the solar metallicity~\cite{Woosley:2002zz}. Note that the SN simulations of the solar metallicity progenitor contains the entire remaining mass of 6.36~M$_\odot$. Both SN simulations employ the DD2F-RDF-1.2 hadron--quark hybrid EOS, reviewed in Sect.~\ref{sec:eos}. 

\begin{figure*}
\adjustbox{valign=t}{%
\begin{minipage}[t]{.25\linewidth}
\caption{Post-bounce evolution of the mass accretion rate, $\dot{M}$, sampled at the PNS surface, comparing the SN simulations of the 75~M$_\odot$ progenitors of solar metallicity and $10^{-4}$ solar metallicity. The inlay shows a zoom in the later post-bounce evolution for the solar metallicity model.\label{fig:mdot}
}
\end{minipage}}%
\hspace{9mm}
\adjustbox{valign=t}{%
\begin{minipage}[t]{0.75\linewidth}
  \vspace{5mm}
  \begin{subfigure} 
  \centering
  \includegraphics[width=0.9\textwidth]{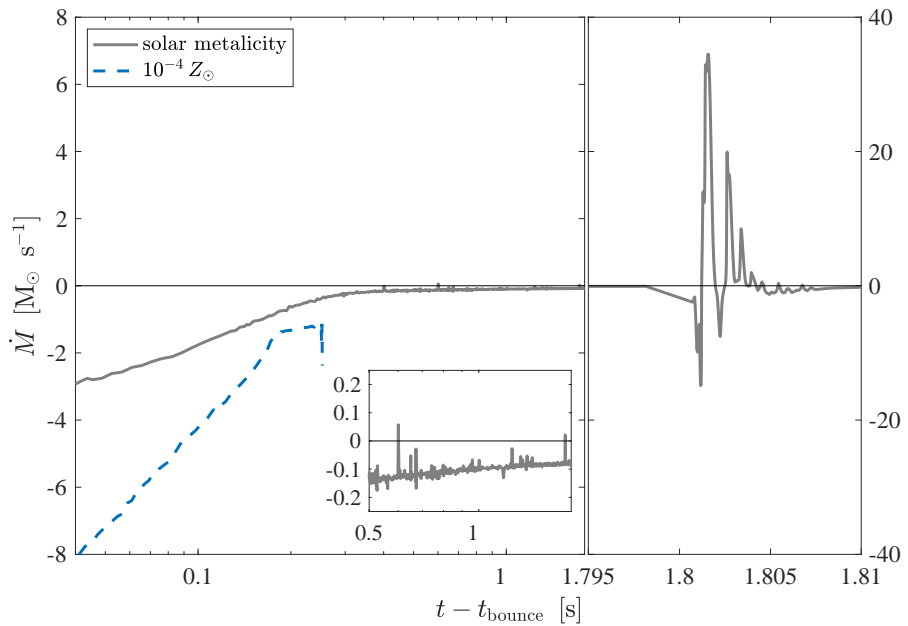}
  \end{subfigure}\par\bigskip
\end{minipage}
}
\end{figure*}

\begin{figure*}
\adjustbox{valign=t}{%
\begin{minipage}[t]{.25\linewidth}
\caption{Evolution of the neutrino luminosities (top panel) and average energies (bottom panel) for all neutrino flavors, $\nu_e$, $\bar\nu_e$ and $nu_x$ collectively denoted for all heavy lepton flavor neutrinos and $\bar\nu_x$ for their antineutrinos respectively, sampled in the co-moving frame of reference at 500~km, for the SN simulation of the low-metallicity 75~M$_\odot$ progenitor with $Z=10^{-4}\times Z_\odot$.\label{fig:h75b_neutrinos}}
\end{minipage}}%
\hspace{9mm}
\adjustbox{valign=t}{%
\begin{minipage}[t]{0.75\linewidth}
  \vspace{14mm}
  \begin{subfigure} 
  \centering
  \includegraphics[width=0.9\textwidth]{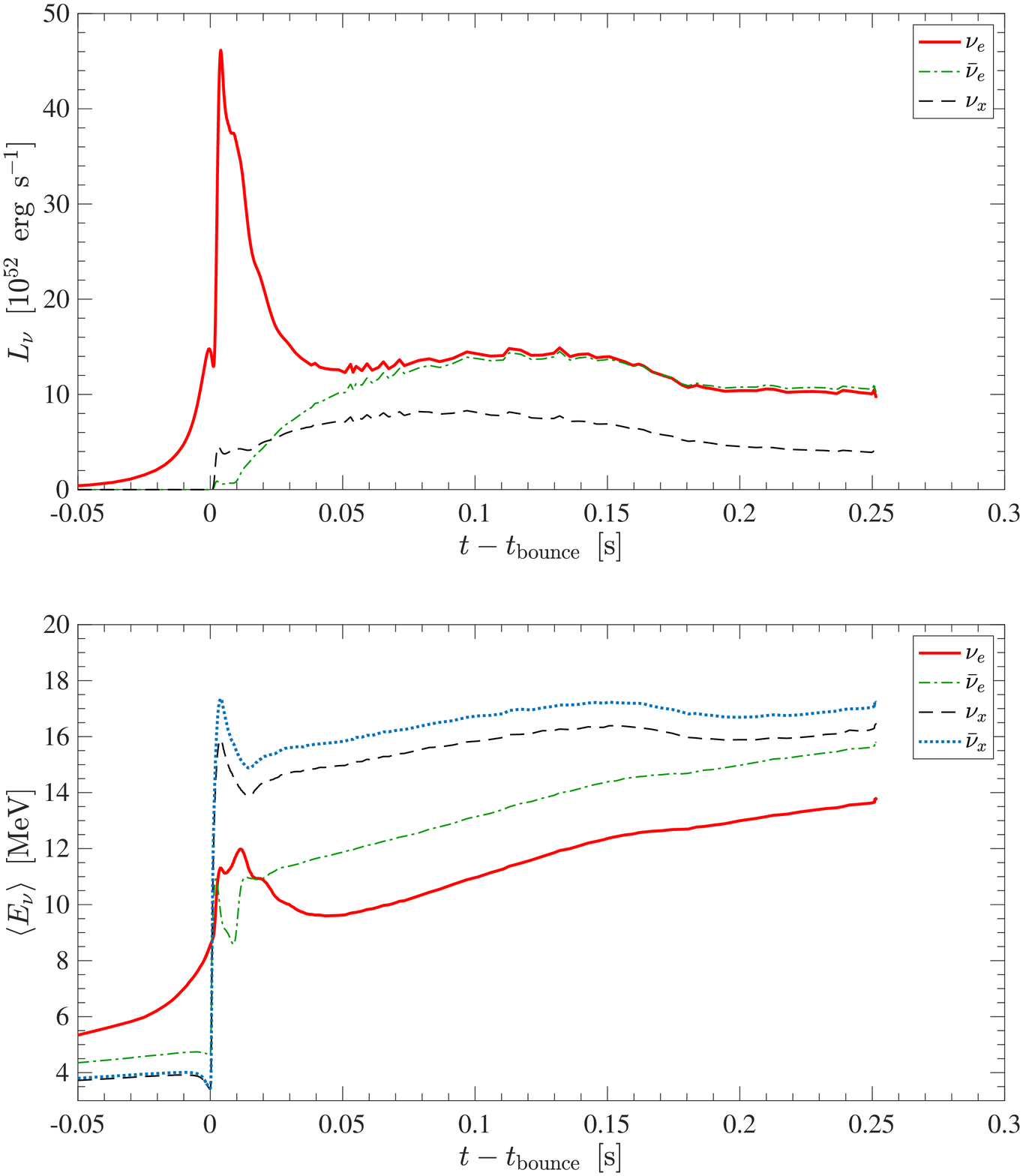}
  \end{subfigure}\par\bigskip
\end{minipage}
}
\end{figure*}

Both SN simulations proceed qualitatively similarly through the stellar core collapse, featuring slightly higher central temperatures for the low-metallicity model as well as lower central density, than the solar-metallicity one, as illustrated in graphs~(a) and (b) in Fig.~\ref{fig:evolution}. However, of particular interest is the post-bounce evolution, during which differences grow substantially between the two models. These, in turn, can be related to the mass accretion rates, shown in Fig~\ref{fig:mdot}, sampled at the PNS surface. The mass accretion rate, $\dot{M}=4\pi r^2 \rho v$, with rest-mass density $\rho$ and velocity $v$ at the PNS surface, is significantly larger for the low-metallicity case as a direct consequence of the higher density of the far more massive iron core and of the extended silicon-sulphur layer above the iron core (see the left panels in Fig~\ref{fig:progenitor}). The mass enclosed inside the PNS  grows very rapidly shortly after core bounce, reaching shortly 2~M$_\odot$, as illustrated in Fig.~\ref{fig:evolution}c. It leads to a rapid rise of the central density and of the central temperature, and in particular to maximum temperatures at the PNS interior, denoted as $T_{\rm max}$, in excess of 30--50~MeV. The very massive PNS gives rise to high neutrino luminosities, on the order of $L_{\nu}\simeq 1-2\times 10^{53}$~erg~s$^{-1}$ (see Fig.~\ref{fig:h75b_neutrinos}) during the evolution after the $\nu_e$-deleptonization burst has been launched, about 50~ms post-bounce. This is nearly one order of magnitude higher than the one obtained in {\em normal} SN simulations after the high-density silicon-sulfur layer has fallen onto the SN shock (c.f. Refs.~\cite{Lentz:2015,Mueller:2015,Mirizzi:2016,Mueller:2017,Kotake:2018,Burrows:2019MNRAS,Bollig:2020arXiv201010506B} and references therein). Exactly this phenomenon has not been the case for the low metallicty model here, for the entire time of the post-bounce evolution the neutrino luminosities are powered by extraordinary high mass accretion rates, being approximated as accretion luminosity for $\nu$ and $\bar\nu_e$ as follows,
\begin{eqnarray}
L_{\nu_e} \propto 2\times 10^{53}~{\rm erg~s}^{-1} 
 \left(\frac{M_{\rm PNS}}{2~{\rm M}_\odot} \right)
\left(\frac{\dot{M}}{4~{\rm M}_{\odot}{/s}} \right)
\left(\frac{10^2~{\rm km}}{R_{\nu}}\right)~. \nonumber \\
\end{eqnarray}
The values for the PNS mass, $M_{\rm PNS}$, is larger than for the solar metallicity case by about one third, and the mass accretion rate is larger my more than a factor of two (see Figs.~\ref{fig:evolution} and \ref{fig:mdot}). The high neutrino luminosities are the reflection of enhanced neutrino heating rates, which, in turn, result in larger neutrinosphere radii and also a systematically larger shock radius, as illustrated in Fig.~\ref{fig:evolution}d. Such phenomenon has been reported based on the SN simulation of a zero-metallicity stellar progenitor with ZAMS mass of 70~M$_\odot$~\cite{Kotake:2018MNRAS}, which, however, eventually lead to a failed SN explosion and the subsequent black hole formation at about 253~ms post bounce for this low-metallicity model. It is associated with the sudden rise of the maximum temperature and central density (see graphs~(a) and (b) in Fig.~\ref{fig:evolution}) and the abrupt cease of the mass accretion rate (see Fig.~\ref{fig:mdot}).

Here, for the low-metallicity 75~M$_\odot$ progenitor, the high mass accretion rate during the early post-bounce evolution results in a critical behavior when the conditions are reached for the hadron--quark phase transition. These are illustrated in Fig.~\ref{fig:evolution}~(b) through thin horizontal lines, showing the critical density for the onset of the phase transition, $\rho_{\rm onset}=2\times\rho_{\rm sat}$, for the corresponding temperature, shown in Fig.~\ref{fig:evolution}~(a), and a representative proton fraction of $Y_p=0.3$. Once the phase transition takes place, the central quark-matter volume fraction starts to rise rapidly for this model and the PNS becomes gravitationally unstable due to the reduced pressure gradient in the co-existence region, between the two stable hadronic and quark matter phases, as was discussed in Sect.~\ref{sec:eos}. The PNS collapse proceeds supersonically. It halts when a sufficient mass at the PNS interior is converted into the quark matter phase, where the EOS stiffens again. As a consequence, a second shock wave forms which propagates to increasingly larger radii where it turns into an accretion front. However, before the accretion front can accelerate along the decreasing density gradient towards the PNS surface, the PNS collapses behind this accretion front. This is related to the fact that the enclosed PNS mass of $M\simeq 2.6$~M$_\odot$ (see graph~(c) of Fig.~\ref{fig:evolution}) exceeds the maximum mass of the DD2F-RDF-1.2 hybrid EOS (see Table~\ref{tab:eos}). This phenomenon has already been reported in Ref.~\cite{Fischer:2018} for the most compact stellar progenitor with ZAMS mass of 25~M$_\odot$ explored in that study. 

\begin{figure*}
\adjustbox{valign=t}{%
\begin{minipage}[t]{.25\linewidth}
\caption{The same as Fig.~\ref{fig:h75a_neutrinos} but for the SN simulations for the solar-metallicity 75~M$_\odot$ progenitor. \label{fig:h75a_neutrinos}}
\end{minipage}}%
\hspace{9mm}
\adjustbox{valign=t}{%
\begin{minipage}[t]{0.75\linewidth}
  \vspace{9mm}
  \begin{subfigure} 
  \centering
  \includegraphics[width=0.9\textwidth]{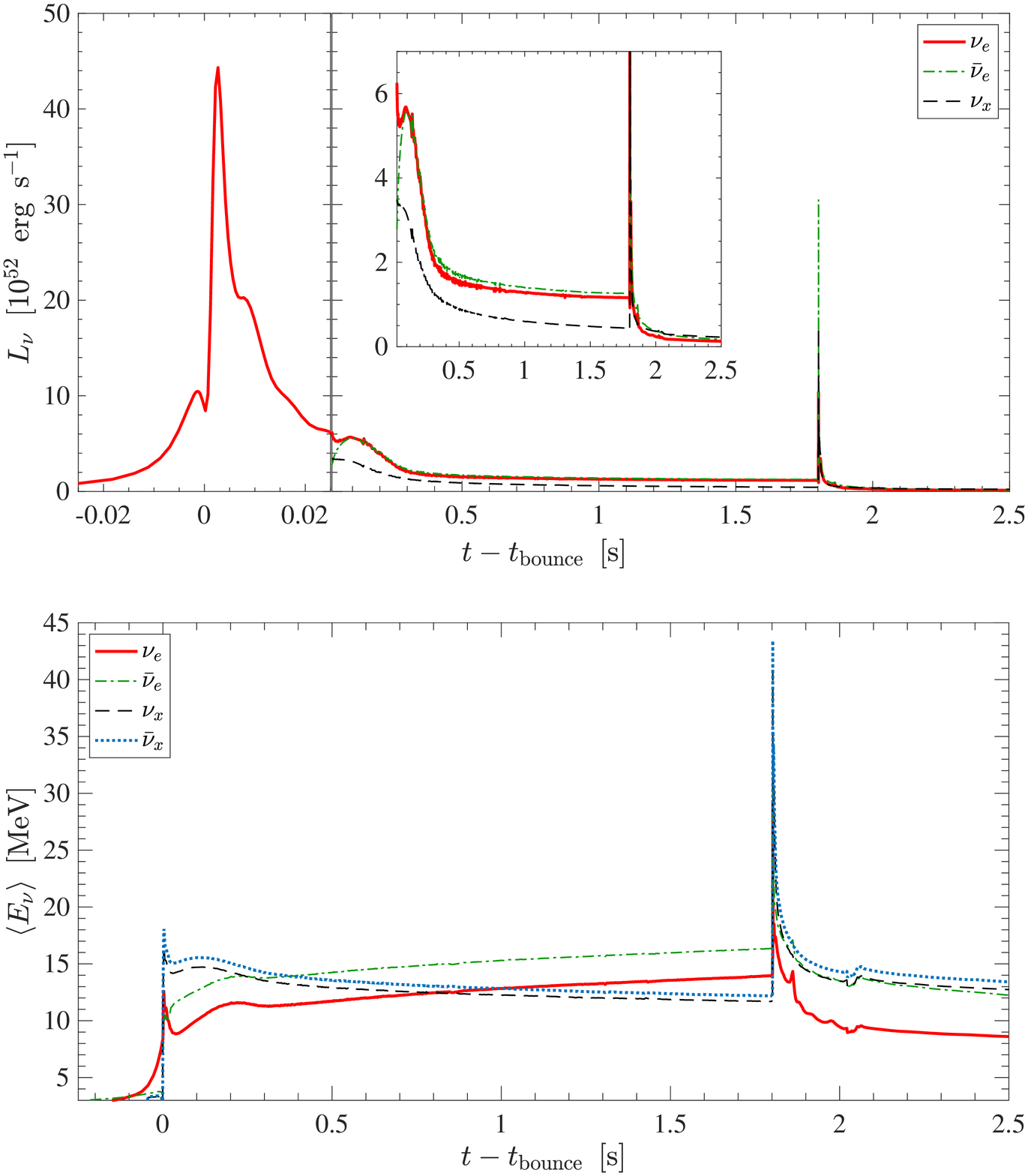}
  \end{subfigure}\par\bigskip
\end{minipage}
}
\end{figure*}

The SN post-bounce evolution proceeds differently for the progenitor of solar metallicity, featuring a substantially lower mass accretion during the entire post-bounce evolution (solid line in Fig.~\ref{fig:mdot}). It results in an extended post-bounce mass accretion phase which lasts for nearly 1.0~second, until the central density reaches a value of about twice saturation density, in comparison to the low metallicity case when the same central density is reached already at about 253~ms post bounce (see graph~(b) in Fig.~\ref{fig:evolution}). The low mass accretion rate for the solar metallicity case, with $\dot{M}\leq 0.1$~M$_\odot$~s$^{-1}$ already after about 500~ms post bounce, leads to a long timescale for the PNS mass to grow. In fact the PNS mass never reaches the maximum mass of the DD2F-RDF-1.2 hybrid EOS, it stays around $M~{\rm PNS}\simeq 1.8$~M$_\odot$ (see graph~(c) in Fig.~\ref{fig:evolution}). Moreover, the SN shock contracts below $R_{\rm shock}<100-80$~km once the silicon-sulfur layer has been fallen onto the shock, and the neutrino luminosities decrease to values on the order of $L_{\nu_x}\simeq 1\times 10^{51}$~erg~s$^{-1}$ ($\nu_x$ denote collectively all heavy lepton flavors, here $\nu_x=\nu_\mu$) and $L_{(\bar\nu_e)\nu_e}\simeq 2\times 10^{51}$~erg~s$^{-1}$ (see Fig.~\ref{fig:h75a_neutrinos}).

\begin{figure*}
\adjustbox{valign=t}{%
\begin{minipage}[t]{.275\linewidth}
\caption{Evolution of the SN simulation of the 75~M$_\odot$ progenitor, of solar metallicity, in the $T-\rho$ phase diagram where the color-coding is due to the electron fraction, $Y_e$. Indications for the phase transition are set as onset of the quark matter phase, $\rho_{\rm onset}$, and reaching the pure quark matter phase, $\rho_{\rm final}$. The evolution of the central temperature and density are also shown (thick solid black line), together with curves of constant entropy per particle for selected values of $s=1-7~k_{\rm B}$ at fixed proton fraction of $Y_p=0.3$ (solid green lines).\label{fig:phasediagram}}
\end{minipage}}%
\hspace{9mm}
\adjustbox{valign=t}{%
\begin{minipage}[t]{0.725\linewidth}
  \vspace{5mm}
  \begin{subfigure} 
  \centering
  \includegraphics[width=0.9\textwidth]{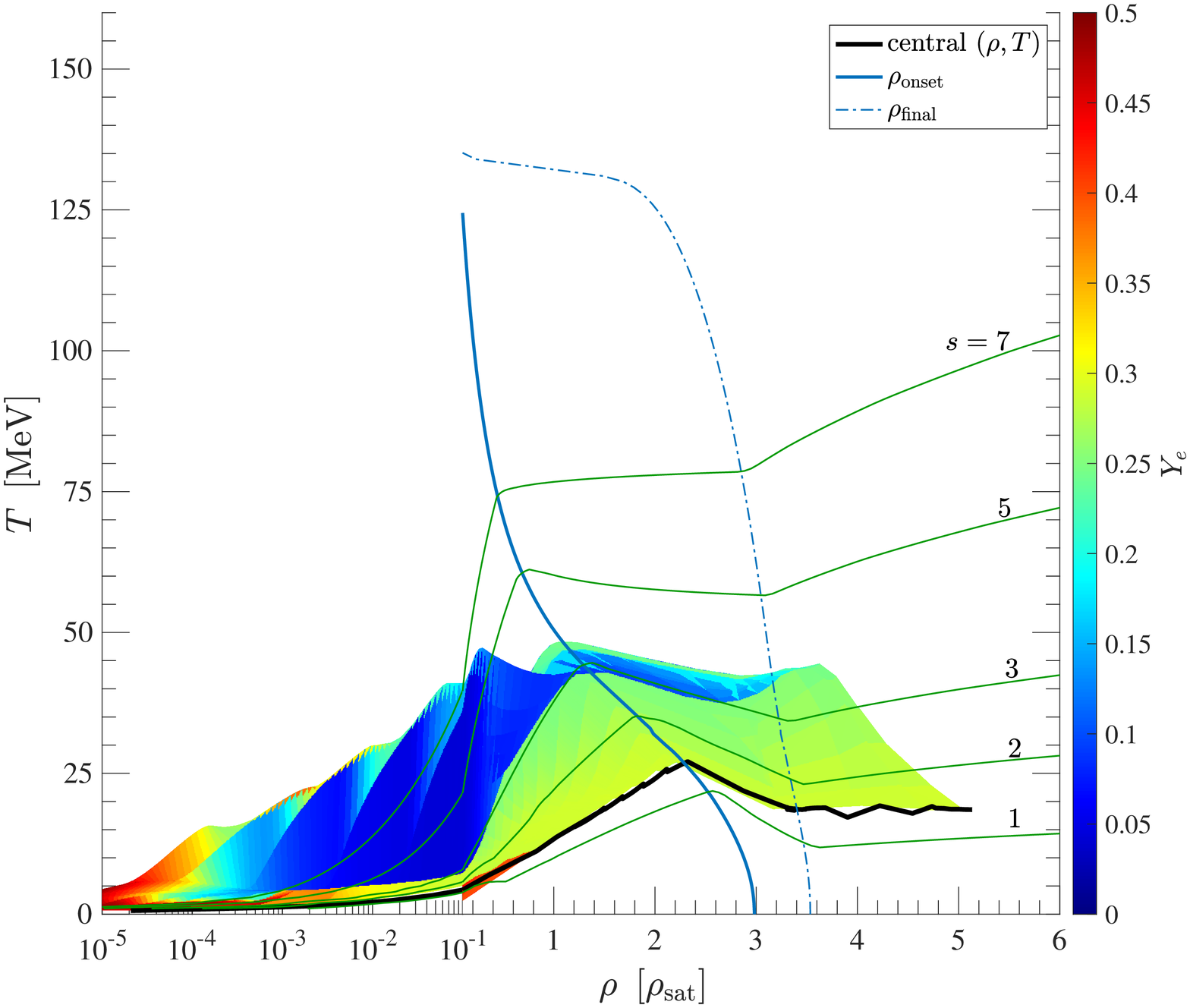}
  \end{subfigure}\par\bigskip
\end{minipage}
}
\end{figure*}

Subsequently, the conditions for the hadron--quark phase transition are reached only after about 1.795~s post bounce for the solar-metallicity case, corresponding to an onset density for the phase transition of $\rho_{\rm onset}=5.6\times 10^{14}$~g~cm$^{-3}$, indicated by the thin horizontal line in graph~(b) of Fig.~\ref{fig:evolution}, for the central temperature of $T_{\rm central}\simeq 30$~MeV, shown in graph~(a). Note that also for this model, maximum temperatures reach as high as $T_{\rm max}\simeq 40-50$~MeV. However, unlike in the case for the low-metallicity model discussed above, the second shock wave, which is associated with the PNS collapse and transition into the hybrid branch, expands to continuously larger radii along the decreasing density at the PNS surface while the PNS remains in a metastable stable. The enclosed mass is well below the maximum mass of the hybrid EOS. During the PNS collapse, the central density rises slightly above $\rho_{\rm central}=5\times\rho_{\rm sat}$, and settles back to a value of about $\rho_{\rm central}=4.2\times\rho_{\rm sat}$ once the second shock wave accelerates. The latter quickly reaches the location of the bounce shock, at a radius of about 50~km (see graph~(d) in Fig.~\ref{fig:evolution}), taking over the bounce shock and expanding continuously to increasingly larger radii. It reaches a radius of 1000~km about 10~ms after its formation  (thin red dash-dotted line in graph~(d) of Fig.~\ref{fig:evolution}). It defines the onset of the SN explosion triggered by the PNS collapse due to the hadron--quark phase transition, indicated by the sudden rise of the diagnostic energy, $E_{\rm diagnostic}$, shown in graph~(e) of Fig.~\ref{fig:evolution},
\begin{equation}
E_{\rm diagnostic}(t) = -\int_{\rm M}^{a_{\rm cut}} da\,E_{\rm specific}(t,a)~,
\end{equation}
which is the baryon mass integration, $da$, of the total specific energy, $E_{\rm specific}$, integrated from the stellar surface, M, towards the center. The specific energy is given as follows, 
\begin{equation}
E_{\rm specific}(t,a) = \Gamma e + \frac{2}{\Gamma+1}\left(\frac{u^2}{2}-\frac{m}{r}\right)~,
\end{equation}
with metric function, $\Gamma=\sqrt{1+m/r-u}$, specific internal energy, $e$, velocity, $u$, and the gravitational mass, $m$. The integration is performed until the mass cut, $a_{\rm cut}$. The latter defines the mass coordinate inside of which material is not ejected, which is determined at the end of the simulation time of about 12~s post bounce. Note that it includes the neutrino-driven wind phase. Note further that the diagnostic energy contains the gravitational binding energy of the stellar envelop, since the entire progenitor mass is included in the SN simulations. 

The diagnostic energy evolves during the SN simulations (see graph~(e) in Fig.~\ref{fig:evolution}). It is often used as indicator for a successful SN explosion when $E_{\rm diagnostic}>0$, referring to the SN explosion energy which is the kinetic energy of the ejecta which should match the asymptotic value obtained for $E_{\rm diagnostic}$ (see the inlay of graph~(e) in Fig.~\ref{fig:evolution}).

The entire SN evolution for the solar-metallicity model is illustrated in the temperature-density phase diagram, Fig.~\ref{fig:phasediagram}, with the color coding of the electron fraction. The evolution of the central values is given as thick black line. It is interesting to note that the central evolution across the phase co-existence, marked by the blue thick solid (onset density) and thin dash-dotted lines (reaching the pure quark matter phase), proceeds along curves of constant entropy per particle of about $s=1.5-3~k_{\rm B}$, which indicates the adiabatic collapse of the PNS. As the entropy is conserved, i.e. neutrinos cannot leave the hot and dense PNS interior, the temperature decreases during the phase transition slightly. This phenomenon has already been realized in Ref.~\cite{Fischer:2018}. Consequently, when considering the stability of hybrid matter PNS, one should take into account the maximum mass of compact stars at finite entropy per particle (see Table~\ref{tab:eos}). It is interesting to note that the SN evolution in the phase diagram is similar as those of binary neutron star mergers, which consider the hadron quark phase transition~\cite{Bauswein:2019PhRvL,Most:2019PhRvL,Blacker:2020}, however, reaching somewhat higher temperatures.

The rapid SN shock expansion of the second shock across the neutrinospheres releases an outburst of neutrinos of all flavors, however, dominated by $\bar\nu_e$ and heavy lepton flavor neutrinos (see Fig.~\ref{fig:h75a_neutrinos}). This burst lasts for only a few tens of a second. The associated luminosities and average neutrino energies rise to values of several times $10^{53}$~erg~s$^{-1}$ and 40~MeV, respectively. It has been pointed out that this non-standard neutrino signal---not present in neutrino-driven SN simulation models---is a smoking-gun signature for a sufficiently strong first-order phase transition. It is observable at the present generation of neutrino detectors, e.g., Super-Kamiokande~\cite{Fischer:2018}.

\section{Summary and conclusions}
\label{sec:sum}
The present article reports about results of two core-collapse SN simulations of very massive progenitor stars with the same ZAMS mass of 75~M$_\odot$ but different metallicities. This important aspect was not explored in previous studies. The metallicity has important consequences for the mass loss during the stellar evolution; high(low) metallicity results in high(low) mass loss rates and consequently in high(low) compactness of the nascent iron core at the onset of stellar core collapse. This, in turn, results in substantial differences during the SN post-bounce evolution, featuring low(high) mass accretion rates. This study considers an EOS with a first-order phase transition from normal nuclear matter to the quark--gluon plasma at high baryon density. In the subsequent SN simulations, there are two competing time-scales: {\em (i)}~the growth of the central density and temperature to reach the critical conditions for the onset of the phase transition, and {\em (ii)} the growth of the mass enclosed inside the PNS. 
Both are determined by the mass accretion rate of the still collapsing stellar core onto the bounce shock and by the compression behavior of the PNS, which, in turn is determined by the high-density behavior of the nuclear EOS. Here, the previous finding are confirmed that a high mass accretion rate may result in a massive PNS at the onset of the hadron--quark phase transition, exceeding the maximum mass of the EOS. Only when the timescale for the PNS compression, i.e. the timescale associated with the rising central density and temperature, exceeding the growth timescale of the PNS mass, the nascent PNS will be in a meta-stable state after the phase transition. The formation of a second shock wave determines the onset of the SN explosion. In the present work, this is the case for the progenitor of solar metallicity only. Contrary, for the low-metallicity model, the enclosed mass exceeds the maximum hybrid EOS mass and hence a black hole forms. The SN explosion of the solar-metallicity model confirms the unique observable signature in the neutrino signal. 

These findings have important consequences for the dependence on metallicity of this class of massive star explosion, driven by a sufficiently strong hadron--quark phase transition. While the first generation of stars had zero metal content, the metal enrichment is considered to rise continuously during the evolving universe. Such considerations lead to the age--metallicity relation~\cite{Hirai:2015}. Here, a metallicity of $10^{-4}$ of the solar value is considered as representative example of extremely metal poor stars~\cite{Umeda:2003ut,Tominaga:2007pb}. Observations of metal poor stars with surface abundances enriched with heavy $r$-process elements, such as strontium and europium, have been observed frequently~\cite{Sneden:2008,Roederer:2012ApJS,Roederer:2012ApJ,Frebel:2018,Cowan:2021RvMP}. It is yet incompletely understood which astrophysical sites can account for these observations, especially at metalicities of $Z=10^{-4}\times Z_\odot$ and below. Specifically, whether binary neutron star mergers can account therefore is presently unclear~\cite{Argast:2004A&A,Wehmeyer:2015,Wehmeyer:2019}. On the other hand, even though SN explosions could have enriched the galaxy at low metallicity, {\em canonical}, neutrino-driven core-collapse SN explosions cannot account for the production of heavy neutron-capture elements heavier than molybdenum, with atomic number of $Z = 42$~\cite{MartinezPinedo:2014}. The reason for this is the absence of sufficient neutron rich ejecta. It points to rare and, perhaps speculative sites for the $r$ process associated with massive star explosions, such as magneto-rotational driven jet-like explosions~\cite{Winteler:2012,Moesta:2014,Moesta:2015,Kotake:2020}. In addition, SN driven by the hadron--quark phase transition have been suggested as possible but rare $r$-process site~\cite{Fischer:2020b}, with an abundance pattern featuring an overproduction of nuclei in the vicinity of the second $r$-process peak in the region with nuclear mass numbers $140<A<180$. It was suggested that this site may operate at low metallicity and hence account for the observed metal-poor star data. However, the findings of the present study points to the absence of this class of massive star explosions driven by the hadron-qaurk phase transition at a metallicity of $Z\leq 10^{-4}\times Z_\odot$.

Zero- and low-metalicty stellar evolution calculations predict qualitatively the same evolutionary pattern for massive stars with ZAMS masses $>30$~M$_\odot$~\cite{Woosley:2002zz,Umeda:2003ut,Umeda:2007wk,Tominaga:2007pb}, namely an extended high-density silicon-sulfur layer above the iron core due to negligible mass loss. Consequently, all these massive stars would belong to the failed SN branch with the failed SN shock revival and hence the formation of black holes. It excludes SN explosions of massive stars driven by the hadron--quark phase transition as $r$-process site operating at low metallicity. At which value of the increasing metallicity during the galactic evolution they might have started to contribute remains to be explored. \\ \\
{\bf Acknowledgements}~The author acknowledges support from the Polish National Science Center (NCN) under Grant No. 2019/ 33/B/ST9/03059. The SN simulations were performed at the Wroclaw Center for Scientific Computing and Networking (WCSS) in Wroclaw (Poland).  \\ \\
{\bf Data Availability Statement}~This manuscript has associated data in a data repository.  The data associated with the equation of state and the supernova simulations are available from the corresponding author on reasonable request. \\ \\
{\bf Open Access}~This article is licensed under a Creative Commons Attribution 4.0 International License, which permits use, sharing, adaptation, distribution and reproduction in any medium or format, as long as you give appropriate credit to the original author(s) and the source, provide a link to the Creative Commons licence, and indicate if changes were made. The images or other third party material in this article are included in the article’s Creative Commons licence, unless indicated otherwise in a credit line to the material. If material is not included in the article’s Creative Commons licence and your intended use is not permitted by statutory regulation or exceeds the permitted use, you will need to obtain permission directly from the copyright holder. To view a  copy of this licence, visit http://creativecommons.org/licenses/by/4.0.

\bibliographystyle{ieeetr}


\end{document}